\documentclass[12pt]{iopart}

\usepackage{graphicx}
\usepackage{etoolbox}
\usepackage{iopams,wrapfig}
\usepackage{xspace}
\usepackage{subfig}
\usepackage{amsmath}
\usepackage[bottom]{footmisc}
\usepackage{hyperref}
\usepackage{graphicx}
\usepackage{cite}
\usepackage{csquotes}

\usepackage{booktabs} 

\usepackage{xcolor} 
\usepackage{tabularx} 

\begin{document}

\title[]{Benchmarking the Non-flow Contributions to the Elliptic Flow Parameter ($v_{_2}$) in Proton-Proton Collisions}

\author{M. I. Abdulhamid$^{1,2,}$\textsuperscript{\footnotemark[1]}, A. M. Hamed$^{1}$, E. A. Osama$^{1}$, M. Rateb$^{1}$, N. K. Sadoun$^{1}$ and F. H. Sawy$^{1}$}

\address{$^{1}$Department of Physics, The American University in Cairo, New Cairo 11835, Egypt}
\address{$^{2}$Department of Physics, Faculty of Science, Tanta University, Tanta 31527, Egypt}

\footnotetext[1]{Corresponding author: \href{mailto:muhammad.ibrahim@science.tanta.edu.eg}{muhammad.ibrahim@science.tanta.edu.eg}} 

\vspace{10pt}

\begin{abstract}
This manuscript reports on the elliptic flow parameter, $v_{_2}$, in proton-proton (p-p) collisions using PYTHIA8 event generator simulations. The typical Event Plane ($EP$) method $v_{_2}$($EP$) as the one used for the real data analysis has been adopted to measure the $v_{_2}$ of particles composed of different quark flavors, and produced at mid-pseudorapidity $|\eta| \le 1$ at center-of-mass energy $\sqrt{s}$ = 200 GeV and $\sqrt{s}$ = 13 TeV. The event plane has been constructed using the soft charged particles with transverse momentum ($p{_{_T}} < $ 2 GeV/c) produced in various $\eta$ ranges. The pseudorapidity gap technique ($\Delta\eta$) between the particle of interest and the soft charged particles contributed to the event plane constructions has been utilized to benchmark the non-flow contributions. The $v_{_2}^{\pi^{^\pm,}\pi^{^0}}$, $v_{_2}^{J/\psi,\Upsilon, K, D, B}$, and the $v_{_2}^{\gamma_{dir}}$ show similar pattern dependence on the transverse momentum for the same $\Delta\eta$, at both $\sqrt{s}$ = 200 GeV and $\sqrt{s}$ = 13 TeV. At each center-of-mass energy, the measured $v_{_2}$ shows obvious dependence on the quark flavor contents where $v_{_2}^{\gamma_{dir}}$ $\le$ $v_{_2}^{J/\psi, \Upsilon, K, D, B}$ $\le$ $v_{_2}^{\pi^{^\pm,}\pi^{^0}}$ for similar $\Delta\eta$ gap. The measured $v_{_2}$ shows slightly higher values at smaller $\sqrt{s}$ for particles of similar values of $\Delta\eta$ gaps, particularly at small $\Delta\eta$ gap. The measured $v_{_2}$ for all particles shows systematic decreases with increasing the $\Delta\eta$ gap as expected from the previous experimental results. Because PYTHIA8 simulations do not incorporate final state interactions, the $v_{_2}(EP)$ values reported here primarily reflect the non-flow contributions and its dependence on the quark contents, pseudorapidity gap as a function on $p_{T}$ at each center of mass-energy. These contributions should be carefully subtracted from the signal in real data analysis that employs the same event plane determination algorithm.
\end{abstract}

\textbf{keywords:} $v_{_2}$, Quark-Gluon Plasma, Non-flow contributions, Reaction plane 
\newpage
\section{Introduction}
\noindent
The hydrodynamic flow patterns have been studied extensively in the system created by heavy-ion collisions, as they provide information about the dynamical evolution of the formed system and its transport coefficients \cite{lim2019examination,schenke2011flow}. The collective hydrodynamic behavior of the created medium has been assessed using many correlations, e.g., a few-particle correlation and the correlations of the emitted particles concerning the reaction plane orientation. These studies have played an important role in constraining the created medium properties and its dynamics\cite{lim2019examination,luzum2011flow}. Recent studies have indicated a prospective examination of the collective and elliptic flow effects in small-sized systems such as proton-proton (p-p) collisions\cite{lim2019examination}. In contrast to the freely streaming particles where there are no final state interactions; the azimuthal distribution of the produced particles in heavy-ion collisions is expected to be sensitive to the initial geometric overlap of the colliding nuclei, resulting in anisotropic azimuthal distributions with respect to the event plane. The conventional approach for quantifying azimuthal elliptic anisotropy involves expressing particle azimuthal distributions through a Fourier series expansion \cite{voloshin1996flow,puri2020advances}:\\
\begin{equation}
\frac{dN}{d\phi}=\frac{N}{2\pi}\left[1+2\sum_{n=1}^{\infty}{v_{_n}(p_{_T},y)\cos{[n\left(\phi_{p_{_T}}-\psi_{_{EP}}\right)}]}\right]
\label{eq1}
\end{equation}
where $\phi$ is the azimuthal angle and \textit{N} is the total number of produced particles. The $\frac{dN}{d\phi}$ ratio represents the differential yield of particles as a function of the azimuthal angle. $\phi_{p_{_T}}$ is the azimuthal angle of the produced particle with a certain value of transverse momentum ($p_{_T}$). $\psi_{_{EP}}$ is the angle of the event plane and $v_{_n}(p_{_T},y)$ as a function of $p_{_T}$ and rapidity ($y$) denotes the coefficient of the $n^{th}$ harmonic, which refers to a specific type of flow. The flow is directed when $n=1$. The $2^{nd}$ harmonic, $n=2$, includes the parameter $v_{_2}$, which denotes the \enquote{elliptic flow} that strongly depends on the pseudorapidity ($\eta$) and transverse momentum $(p_{_T})$ \cite{luzum2011flow}. The elliptic flow $v_2$ is given by:
\begin{equation}
v_{_2}(p_{_T},\eta) = \langle e^{2i(\phi_{p_{_T}}-\psi_{_{EP}})} \rangle = \langle \cos[2(\phi_{p_{_T}}-\psi_{_{EP}})]\rangle 
\label{TSP}
\end{equation}
where the angular brackets $\langle...\rangle$ denote the statistical averaging over particles and events. The correlation of flow is given by eq.\ref{eq1}, while any additional contribution will be considered as a \enquote{non-flow} \cite{luzum2011flow}. \\

\noindent
The event plane angle ($\psi_{_{EP}}$) is a critical experimental observable because it allows for the calculation of flow coefficients. It is conventionally determined from the azimuthal distribution of the produced soft charged particles produced in specific pseudorapidity in a given event, and it may differ from the true reaction plane angle due to the finite number of particles, fluctuations, and measurement errors. On the other hand, $\psi_{_{RP}}$ is a theoretical concept that defines the true plane of symmetry in the collision geometry and it cannot be measured directly. Typically, it refers to the azimuthal angle of the plane defined by the beam axis and the impact parameter vector. The reaction plane angle is commonly determined by the initial geometry of the collision and it is important in understanding the initial spatial anisotropy of the collision, which gives rise to the elliptic flow \cite{voloshin2010collective}.\\

\noindent
One of the findings of heavy-ion collisions is that the produced particles demonstrate azimuthal anisotropy in the transverse plane. This anisotropy is a result of the initial geometric anisotropy of the collision as evident from the overlap region of the two colliding nuclei being almond-shaped according to kinematic models. Due to the fact that the produced particles are not freely streaming, this geometric anisotropy is then reflected in a momentum azimuthal anisotropy distribution, which is expressed mathematically by the Fourier expansion shown in eq.\ref{eq1}. The parameter $v_{_2}$, which measures elliptic flow, has been widely studied and is often used as a measure of the momentum azimuthal anisotropy concerning the reaction plane ($RP$). Quantifying $v_{_2}$ is important in studying the thermalization of the hypothesized Quark-Gluon Plasma (QGP) by quantifying the collective flow as well as obtaining information about the equation of state for the strongly interacting medium formed in the collisions. As evident in eq.\ref{TSP}, to calculate $v_{_2}$, one needs to determine the azimuthal angle of the event plane ($\psi_{_{EP}}$), which serves as a proxy for the true reaction plane ($\psi_{_{RP}}$). The measured values of $v_{_2}$ serve as a direct indicator for the degree of collectivity in the medium, such as the possible formation of a thermalized QGP constraining its parameters; e.g. viscosity and entropy.\\

\noindent
As the true reaction plane angle ($\psi_{_{RP}}$) is difficult to determine experimentally, the Event Plane ($EP$) method is widely used as a practical approach to estimate it. The event plane angle ($\psi_{_{EP}}$) can be calculated using the following formula \cite{puri2020advances}:
\begin{equation}
\psi_{{EP}} = \frac{1}{2}\tan^{-1}{\left(\frac{\sum_{i}\sin{(2\phi_i)}}{\sum_{i}\cos{(2\phi_i)}}\right)}
\label{eq3}
\end{equation}\

\noindent
where $\phi_{i}$ is the azimuthal angle of the $i^{th}$ particle, and the sum runs over all soft particles ($p_{_T} < 2$ GeV/c) produced within certain pseudorapidity range; except the particle of interest to eliminate the auto-correlation \cite{zhu2005elliptic,poskanzer1998methods}. A drawback of this method is that energetic jets produced in the collision can bias the event plane angle determination due to the jet fragmentation, and potentially aligning it with the jet direction. This bias introduces a non-flow related anisotropic component to the elliptic flow ($v_{_2}$), which must be accounted for to accurately isolate the true flow component.\\

\noindent
The eccentricity ($\varepsilon$) with the formula mentioned in eq.\ref{eccentricity} \cite{puri2020advances,adcox2005formation,snellings2011elliptic,luzum2009viscous} is a measure of how much an ellipse deviates from a perfect circle. For heavy-ion collisions, the overlap region of the colliding nuclei is typically modeled as an ellipse, and the eccentricity is related to the geometry of this ellipse. It provides a comprehensive account of the spatial irregularities within the region of the initial reaction.
\begin{equation}
\varepsilon=\frac{\langle Y^2 - X^2\rangle}{\langle  Y^2 + X^2 \rangle}.
\label{eccentricity}
\end{equation}
In the previous equation, $X$ and $Y$ denote the spatial coordinates of the nucleons in the overlap region of the two colliding nuclei. $X$ is the horizontal axis, and $Y$ is the vertical axis in the transverse plane (i.e. perpendicular to the beam direction). The brackets indicate the averaging over all nucleons.\\

\noindent
In terms of the Impact Parameter ($IP$), the $\varepsilon$ value fluctuates from one event to another \cite{puri2020advances}. The value of $IP$ can be calculated by eq.\ref{IP} and it is used to determine the centrality of the collision. Eccentricity shapes the elliptic flow of the fireball created in heavy-ion collisions in terms of its expansion \cite{li2010re,bozek2011elliptic,ollitrault1992anisotropy}. There is a correlation between $v_2$, representing the final state, and $\varepsilon$ in the plane perpendicular to the beam axis, signifying the initial state \cite{adil2006eccentricity};
\begin{equation}
    IP = \sqrt{X_{_{CM}}^2+Y_{_{CM}}^2}
    \label{IP}
\end{equation}
\noindent
where $X_{_{CM}}^2$ and $Y_{_{CM}}^2$ represent the magnitudes of IP in the horizontal and vertical directions, respectively, from the center-of-mass of the overlap region in the transverse plane. If $IP = 0$, the collision is central, while it is considered peripheral (i.e. indicating a smaller overlap region between the colliding nuclei) as $IP > 0$.\\

\noindent
The first calculation of $v_2$ for p-p collision at $\sqrt{s}=7$ TeV is given in \cite{bozek2011elliptic}. In a study for p-p collisions at $13$ TeV \cite{cunqueiro2010nuclear}, it is found that the value of $v_2$ is directly proportional to the eccentricity and impact parameter. At the same time, it is inversely proportional to the overlapping area ($S$), where $S=\pi\sqrt{\langle X_{_{CM}}^2 \rangle \langle Y_{_{CM}}^2 \rangle}$. $\langle X_{_{CM}}^2 \rangle$ and $\langle Y_{_{CM}}^2 \rangle$ denote the averages in the horizontal and vertical directions, respectively, and charged particle distribution at central rapidity ($\frac{dN}{dy}$). The ratio $v_2/\varepsilon$ as a function of multiplicity per unit rapidity divided by the overlapping area $\frac{dN/dy}{S}$ can be used as a tool for identifying the elliptic flow \cite{cunqueiro2010nuclear,bautista2009particle}. Several studies have been performed to examine the elliptic flow of different mesons. For the D meson, the directed ($v_1$), elliptic ($v_2$), and triangular ($v_3$) flows were studied at $\sqrt{S_{_{NN}}}= 5.02$ TeV of Pb-Pb collision in the ALICE experiment \cite{trogolo2021directed} where the triangular flow defines $\langle cos(3(\phi - \psi_{3}))\rangle$ and $\psi_{3}$ is the minor axis of the triangular participant. In this study, the values of $v_2$ and $v_3$ were compared for pions. At the same center-of-mass energy and collided nuclei, the elliptic flow of muons decay from the charmed and bottom hadrons at different centrality ranges 0-10\% and 40-60\% were contrasted with theoretical calculations, providing more information about the interaction of the previously mentioned particles with hot-dense matter \cite{hu2020production}, also the p-p collision is examined at $\sqrt{s}$ = 13 TeV. The results show that $v_2$ was not zero for the Pb-Pb central and semi-peripheral collisions. The elliptic flow of charmed mesons from the Pb-Pb collision at $\sqrt{S_{_{NN}}}= 5.02$ TeV and in mid-centrality (i.e. 30-50\%) in the ALICE experiment is reported \cite{acharya2018d}. The $v_2$ reconstruction of charmed mesons was at mid-rapidity lower than 0.8 and a momentum range of 1-24 GeV/c. Comparing the previous average values of $v_2$ with their correspondents at $\sqrt{S_{_{NN}}}= 2.76$ TeV, they were found to be not identical \cite{abelev2013d}. The $v_2$ of photons at high $p_{_T}$ corresponds to the anisotropy in quark momentum and is measured directly after the collision \cite{kasmaei2020photon}. Conversely, at low $p_{_T}$, it reflects the substantial momentum anisotropy of pions during hadronic release (i.e. later stages). At $p_{_T}$$ \leq$ 1-2 GeV/c supports information about the development of the QGP fireball \cite{chatterjee2006elliptic}. \\

\noindent
The medium-induced radiative energy loss of partons (jet-quenching) has been proposed as the source for the large observed azimuthal anisotropy at high $p_{_T}$ due to the path-length dependence of the parton energy loss \cite{shuryak2002azimuthal}. STAR experiment results show that the magnitude of $v_{2}$ at high $p_{_T}$ is larger than the predicted values by pure jet-quenching models \cite{adams2004azimuthal}. The same results presented no connection between the energy loss and the path length at high $p_{_T}$ through a comparison between correlations of $\gamma_{dir}$ and $\pi^0$ with charged particles \cite{abelev2010parton}. Despite the recent results of PHENIX experiment indicating an in-plane predominance of  $\pi^{0}$ particles over those produced out-of-plane, such observations may align with the path-length dependence of energy loss, as reported in the literature \cite{adare2010azimuthal}. However, the reaction plane determination might remain biased toward the direction of the produced jets.\\

\noindent
This paper aims to illustrate the impact of event plane calculation bias caused by jet fragmentations and to establish a benchmark for future research, facilitating the quantification of necessary corrections. In pursuit of this goal, p-p collisions at $\sqrt{s}$ = 200 GeV and 13 TeV (i.e. corresponding to the Relativistic Heavy Ion Collider (RHIC) and the Large Hadron Collider (LHC) maximum energies) were simulated using (PYTHIA 8.2). The flow coefficient $v_{_2}$ has been calculated for mid-rapidity $|\eta| \le 1$ pions (i.e. ${\pi} ^+$, $\pi^-$ and $\pi^0$), mesons containing heavy quarks (J/$\psi$, $\Upsilon$, the D and B meson families) and direct photons ($\gamma_{dir}$) representing real photons originating directly from the electromagnetic vertex. The $EP$ method is applied to 30 million simulated events at $\sqrt{s}$ = 200 GeV as well as $\sqrt{s}$ = 13 TeV. The $EP$ was evaluated using charged particles with $p_{T} <$ 2 GeV in various $|\eta|$ bins as follow: $|\eta|<1$, $1<|\eta|<2$, $2<|\eta|<3$, $3<|\eta|<4$, $4>|\eta|>5$ and $|\eta|>5$. Therefore, this study has three relevant independent variables that affect the value of $v_{2}$. Those variables are $\sqrt{s}$, quark flavors, and different $|\eta|$ bins, while RHIC data show a large amount of elliptic flow as predicted by the hydrodynamic models at low $p_{_T}$, the results at high $p_{_T}$ deviate strongly from the hydrodynamic predictions as expected \cite{adams2005experimental}. \\

\noindent
The assessment of any residual bias in the determination of the reaction plane, potentially resulting in a positive azimuthal elliptic anisotropy ($v_{2}$) signal would be facilitated by measuring the $v{_2}$ of direct photons. Furthermore, the $v_{2}^{\gamma_{dir}}$ would give additional information to help disentangle the various scenarios of direct photon production through the expected opposite contributions to the $v_{2}$ \cite{zakharov2004radiative,fries2003high,turbide2006azimuthal}, which could help to confirm the observed binary scaling of the direct photon \cite{adler2005centrality}. Since PYTHIA8 does not incorporate final state interactions \cite{sjostrand2015introduction}, any $v_{_2}$ measurement is not due to the collective flow; however, they are solely due to anisotropies introduced because of biases in the adopted event plane method. Therefore, the $v_{_2}$ values reported here should be a benchmark to reduce biases from real data calculations and isolate flow components.
\section{Analysis Details}
\noindent
PYTHIA8 (version 8.2) \cite{sjostrand2015introduction} was used to simulate the p-p collisions while maintaining its default parameters, without being tuned to any of the studied collision energies, $\sqrt{s}$ $=200$ GeV and $13$  TeV. The events were generated for each $\sqrt{s}$ energy with a kinematic region of pseudorapidity $|\eta|< 20$ and full azimuth $|\phi| < \pi$. The hadronic decay channel in PYTHIA8 was turned off, to detect the resonances without the need to identify them via the invariant mass reconstruction as in the real experiments. The total number of events generation at each center-of-mass energy is 15 million, where all produced particles per event are identified and their kinematics ($\eta$, $\phi$, and $p_{_T}$) were recorded. Table \ref{table:1} shows more details for the parameters implemented to generate events of interest.

\begin{table}[htbp!]  
\centering
\caption{Summary of Key PYTHIA8 Parameters Used in the Simulation}
\label{tab:pythia8_params}
\centering 
\begin{tabular}{c c} 
\hline\hline   
\textbf{Parameter Name} & \textbf{Value}   \\ \hline
\texttt{Beams:idA}           & 2212 (proton)      \\
\texttt{Beams:idB}           & 2212 (proton)        \\ 
\texttt{Beams:eCM}           &  200 GeV or 13 TeV  \\ 
\texttt{Number of Events}    & 15 million   \\ 
\texttt{PhaseSpace:pTHatMin} & 5.0 GeV     \\ 
\texttt{HardQCD:all}         & \texttt{on}   \\
\texttt{Charmonium:all}      & \texttt{on}  \\ 
\texttt{Bottomonium:all}     & \texttt{on}   \\ 
\texttt{PromptPhoton:all}    & \texttt{on}  \\
\texttt{Random:setSeed}      & \texttt{on}   \\ 
\texttt{PartonLevel:ISR}     & \texttt{on}  \\ 
\texttt{PartonLevel:FSR}     & \texttt{on} \\ 
\texttt{HadronLevel:Decay}   & \texttt{off} \\ \hline
\end{tabular}
\label{table:1}
\end{table}

\subsection{Event Selection Criteria}
The event selection criteria are crucial for ensuring the quality and relevance of the data analyzed. Table \ref{tab:event_cuts} summarizes the event-level selection cuts applied in this analysis. The selection of $p_{_T}$ and \( \eta \) bins is essential for analyzing flow parameters. The following criteria were used to determine the bin ranges:
\begin{itemize}
\item \textbf{Transverse Momentum ($p_{_T}$) Bins:} The $p_{_T}$ bins were selected based on the typical range of transverse momenta observed in proton-proton collisions at the given center-of-mass energies. At lower $p_{_T}$ values, the collective flow effects are more significant, but for higher $p_{_T}$ values, re-binning has been utilized and the bins are wider to maintain statistical significance, reflecting the reduced number of events in this region \cite{alice2018charm, star2005review}. Similar to high-energy experimental setups, soft charged particles ($p_{_T} < 2$ GeV/c), except the particle of interest, produced within the full azimuthal coverage ($|\phi| < \pi$) were used to determine the event plane per event.\\

\item \textbf{Pseudorapidity ($\eta$) Bins:} The pseudorapidity bins were chosen to cover the full kinematic range accessible by the detector, enabling a comprehensive study of the pseudorapidity gap dependence of the flow parameters \cite{ollitrault1993anisotropy, voloshin2008collective}. To explore the effect of the $\Delta\eta$ gap between the particle of interest and the soft particles used for the event plane reconstructions, the $|\eta| < 20$ range was chosen to cover the mid-pseudorapidity to the forward pseudorapidity. Such coverage ranges in pseudorapidity reach the limit of the zero-degree calorimeter in the typical detector acceptance of high-energy physics experiments. In this analysis, we use the absolute pseudorapidity $|\eta|$ which refers to the magnitude of the individual particles. The event plane ($\psi_{_{EP}}$) is calculated separately according to eq.\ref{eq3} for each pseudorapidity bin using the azimuthal angles of the soft-charged particles within that bin. These ranges are divided into the following bins: $|\eta| < 1$, $1 < |\eta| < 2$, $2 < |\eta| < 3$, $3 < |\eta| < 4$, $4 < |\eta| < 5$, and $5 < |\eta| < 20$. Different pseudorapidity regions were selected for reaction plane determinations to minimize the non-flow contributions as expected when the forward pseudorapidity detectors were used to determine the event plane. 
\end{itemize}

\begin{table}[htp!]  
\caption{Summary for the Event Selection Criteria for PYTHIA8 Simulation}
\label{tab:event_cuts}
\centering 
\begin{tabular}{c c} 
\hline\hline   
\textbf{Criterion} & \textbf{Value}\\ \hline
$\eta$ for all particles & $|\eta| < 20$  \\ 
 $\phi$  range & full (i.e. \(-\pi\) to \(\pi\)) \\ 
$p_{_T}$ for $EP$ soft particles & \( p_{_T} < 2.0 \) GeV/c \\ 
Phase space cutoff & $p_{_T} > 5.0$ GeV/c \\ \hline
\end{tabular}  
\end{table}

\subsection{Generated Data and Quality Assurance}
\begin{figure}[h!]
\centering
\includegraphics[width=170mm]{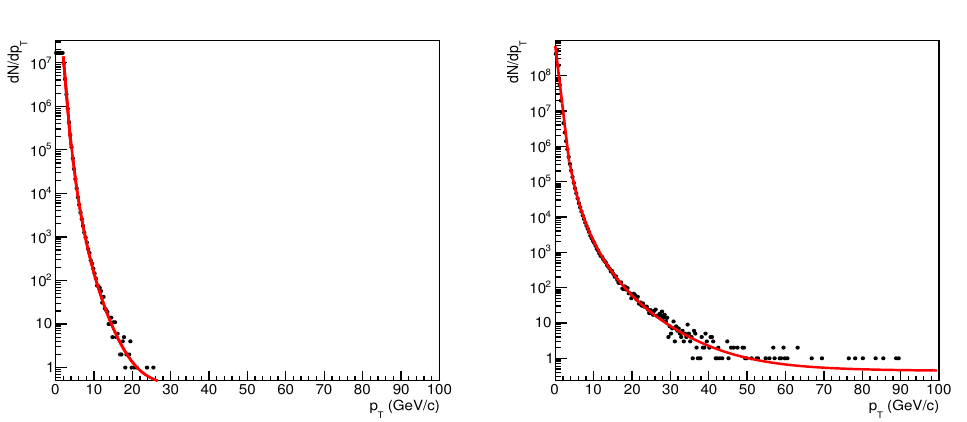}
\caption{Transverse momentum distribution for $\sqrt{s}$ = 200 GeV (left) and $\sqrt{s}$ = 13 TeV for p-p collisions (right)}
\label{fig:1}
\end{figure}

\begin{figure}[h!]
\centering
\includegraphics[width=160mm]{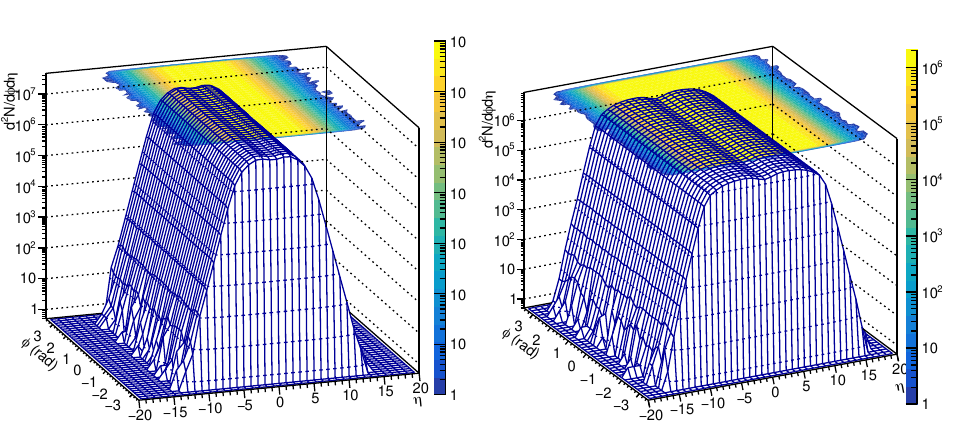}
\caption{The distribution $d^{2}N/d\eta d\phi$, at (left) $\sqrt{s} = 200$ GeV and $\sqrt{s} = 13$ TeV (right)}
\label{fig:2}
\end{figure}

In this study, we rely on generated particle kinematics without incorporating detector effects or reconstruction efficiency corrections that are required due to the limited resolutions of the detectors. Figure \ref{fig:1} shows that particles' distribution also differs as a function of the collision energy. The $dN/dp_{_T}$ distribution at RHIC is better described by the power-law function i.e. $1/{p_{_T}^{n}}$, where $n = 7$ for RHIC energy with $p_{_T}$ in the range $ 0 < p_{_T} < 20$, and $n = 5$ at LHC energy, where $0 < p_{_T} < 40$. The particles' distribution falls more rapidly at RHIC than at LHC. Hence, the distributions at LHC exhibit a higher kinematics reach as expected due to the difference in the available center-of-mass energy at each collider; and accordingly different regions of the probed parton distributions function. 
\\
\noindent
Figure \ref{fig:2} shows the populations of the produced particles in $\eta$ ($|\eta|\le$ 20) as well as in $\phi$ ($|\phi| \le$ 3.14 as $\pi \approx 3.14$) for the two different collision energies (i.e. 200 GeV and 13 TeV). Both plots exhibit approximately uniform patterns within the covered kinematic regions of mid-pseudorapidity, more importantly, the distribution indicates the absence of a bias in the azimuthal direction. This is a crucial factor to ensure in the analysis because any bias in the azimuth would affect the calculation of the azimuthal correlation with the reaction plane ($v_{_2}$). 
\\

\begin{figure}[h!]
\centering
\includegraphics[width=150mm]{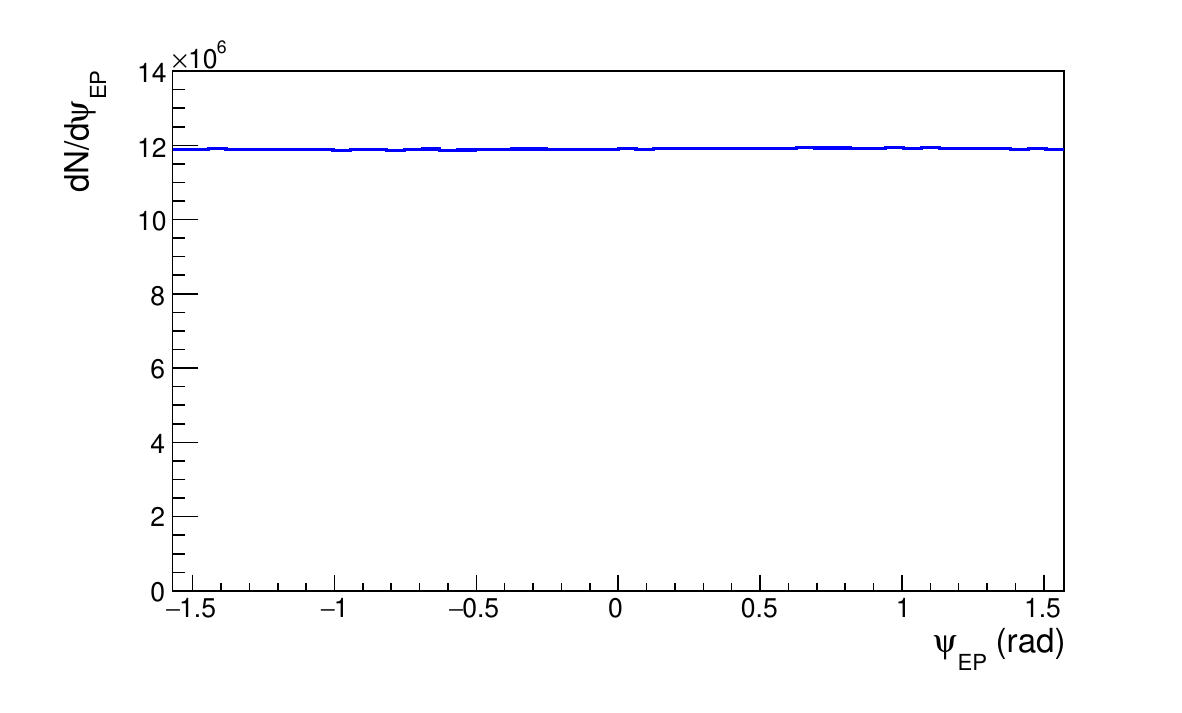}
\caption{ The distribution $dN/d\psi_{_{EP}}$ at $\sqrt{s}$ = $200$ GeV}
\label{fig:3}
\end{figure}

\noindent
Figure \ref{fig:3} depicts the distribution of the calculated event plane angle ($\psi_{_{EP}}$) from eq.\ref{eq3}. This uniformity in the distribution indicates an absence of a bias in the event plane. The calculated values of $\psi_{_{EP}}$ are in the range of $-\pi/2<\psi_{_{EP}}<\pi/2$.

\section{Results}
The relative azimuthal direction with the event plane angle; $v_{_2}$; of pions: $\pi^0$ and $\pi^\pm$, heavy mesons: $J/\psi$, $\Upsilon$, $K^0$, $K_0^{*0}$, $K_0^{*+}$, $K^+$, $D^0$, $D^+$, $D_0^{*0}$, $D_0^{*+}$, $D_s^0$, $B^0$, $B^+$, $B^{*0}$, $B^{*+}$, $B_s^0$ and $B_s^{*0}$, and direct photons: $\gamma_{dir}$ as a function on $p_{_T}$ are presented by figure \ref{fig:4} and \ref{fig:5} at the $\sqrt{s}$ = 200 GeV and $\sqrt{s}$ = 13 TeV respectively. The pattern of $v_{2}$ on $p_{_T}$, and the effect of center-of-mass energy $\sqrt{s}$, pseudorapidity gap $|\eta|$ and the quark flavor contents will be addressed in the following context. 

\subsection{Effect of $\sqrt{s}$ on $v_{_2}$}
A general trend is noticed for the value of $v_{_2}$ which is getting higher at the lower $\sqrt{s}$. The effect is more obvious for lower $|\eta|$ bins since the value of $v_{_2}$ is almost zero for higher $|\eta|$. At lower $\sqrt{s}$ the parton distribution function is dominated by quarks for the same kinematic reach, and hence the quark experiences harder fragmentation than the gluon, therefore, the auto-correlation is expected to be higher. Therefore the $v_2$ goes reciprocally with the $\sqrt{s}$ at fixed $|\eta|$, for the same quark content at similar $p_{_T}$.

\subsection{Effect of $|\eta|$ on ${v_{_2}}$}
Generally, the value of $v_{_2}$ is getting smaller as we go to higher values of the pseudorapidity gap and vanishes at the forward pseudorapidity which indicates the reduction in the non-flow contributions when the forward rapidity regions are used to construct the event plane. The fact that the $v_{_2}$ of direct photons is finite at the mid-pseudorapidity gap indicates the contribution of the jet fragmentation to the reaction plane determination and hence the non-flow component. The direct photon has no near-side yields and accordingly no azimuthal correlations but it is obvious the away side yields still make a sizable effect and cause non-zero $v_{_2}$. The differences in the $v_{_2}$ values of direct photons at low $|\eta|$ between the two center-of-mass energies can be attributed to the different probed regions in the parton distribution functions as mentioned in the previous section. It is also obvious that the light mesons have higher $v_{_2}$ compared to the heavy mesons which can be understood in terms of the different fragmentation functions between the light and heavy mesons for the available kinematics reach. 

\subsection{Effect of quark flavors on $v_{_2}$}
The results show that ${v_2}^\pi({p_{_T}})$ $>$ ${v_2}^H(p_{_T})$ $>$ ${v_{2}}^{\gamma_{dir}}(p_{_T})$, where ${v_2}^\pi$, ${v_2}^H$ and ${v_{2}}^{\gamma_{dir}}$ at low $p_{_T}$ values. On the other hand, there is no obvious trend for higher $p_{_T}$ values due to the insignificant statistics. Regardless of the direct photon $v_{_2}$ which is used as a reference there is obvious quark flavor dependence. For similar kinematics reach in $p_{_T}$, and $|\eta|$ at the same $\sqrt{s}$; the light mesons show higher $v_{_2}$ where the bias in the event plane determination is expected to be higher due to the higher multiplicity of the soft particles compared to the case of heavy mesons. Heavy mesons consume more energy to be created than the lighter ones, leaving less energy for the soft particle creations and their multiplicity. 

\subsection{Effect of $p_{_T}$ on $v_{_2}$}    
At both presented $\sqrt{s}$ in this study, the $v_{_2}$ is finite at low $p_{_T}$ and vanishes at high $p_{_T}$ for all of the particles. Since there are no final state interactions in PYHTIA8, then indeed the observed finite values of $v_{_2}$ at high $p_{_T}$ are due to the medium effect created in the experiment \cite{adams2004azimuthal}. At higher $p_{_T}$ less number of produced particles are produced and accordingly the correlations are eliminated. At such high $p_{_T}$ the eccentricity in the spatial coordinates if any is washed out in the momentum space and particles appear to stream freely. 

\begin{figure}[h!]
	\centering
	\includegraphics[width=170mm]{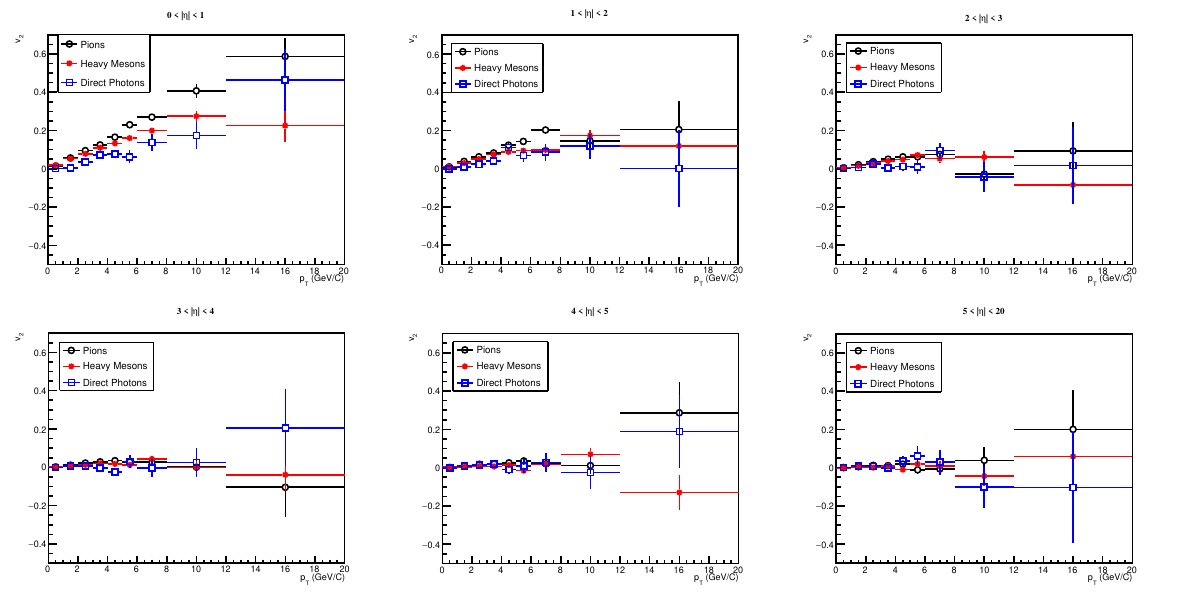}
	\caption {${v_2} ({p_{_T}})$ for pions, heavy mesons and direct photons at different $|\eta|$ bins for $\sqrt{s}$ = 200 GeV with statistical errors}
	\label{fig:4}
\end{figure}
\noindent
\begin{figure}[h!]
	\centering
	\includegraphics[width=170mm]{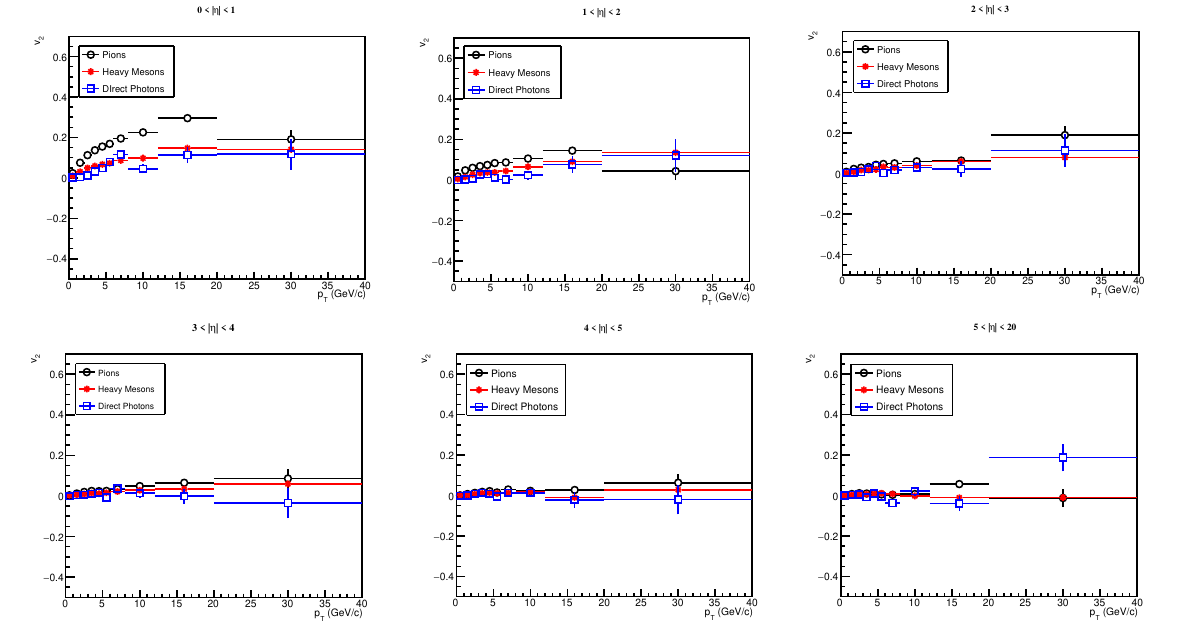}
	\caption {${v_2}({p_{_T}})$ for pions, heavy mesons and direct photons at different $|\eta|$ bins for $\sqrt{s}$ = 13 TeV with statistical errors}
	\label{fig:5}
\end{figure}

\noindent
In Fig.\ref{fig:4} and Fig.\ref{fig:5}, the pions, heavy mesons, and direct photons are always selected within $|\eta| \le 1$, and the $|\eta|$ intervals at the top of each panel represents the pseudorapidity binning for the soft charged particles used for the event plane determination. The higher range of ${p_{_T}}$ and wider binning choice for pp collisions at $\sqrt{s}$ = 13 TeV, compared to $\sqrt{s} = 200$ GeV is for covering the anticipated higher ${p_{_T}}$ ranges for produced particles.

\newpage
\section{Conclusion}
In this study, we performed a detailed analysis of the elliptic flow parameter, $v_2$, of various particles (light mesons, heavy mesons, direct photons) using the conventional event plane method in proton-proton (p-p) collisions at $\sqrt{s}$ = 200 GeV and 13 TeV using PYTHIA8 event generator simulations. The $v_{_2}(p_{_T})$ for the different particles has shown a similar pattern to that measured in the real experiments. This analysis reports the significant dependence of the elliptic flow parameter on the pseudorapidity gap and its importance in eliminating the non-flow contributions. The interplay between the quark flavor, and the center-of-mass energy has a significant impact on the $v_{_2}(p_{_T})$ values. Non-flow contributions to the $v_{2}$ signal, arising from biases in the event plane determination due to jet fragmentations, are shown in this simulation. Since PYTHIA8 does not include final state interactions, the $v{_2}$ values primarily reflect non-flow effects. These contributions should be carefully subtracted in case of experimental data to isolate the true collective flow signal accurately.


\section{References}
\bibliography{bibliography.bib}{}
\bibliographystyle{unsrt}

\end{document}